# ContainerStress: Autonomous Cloud-Node Scoping Framework for Big-Data ML Use Cases

Guang Chao Wang
Oracle Physical Sciences Research Center
Oracle Corporation
San Diego, CA
guang.wang@oracle.com

Kenny Gross
Oracle Physical Sciences Research Center
Oracle Corporation
San Diego, CA
kenny.gross@oracle.com

Akshay Subramaniam
AI Developer Technology Engineering
NVIDIA Corporation
Santa Clara, CA
asubramaniam@nvidia.com

*Abstract*—Deploying big-data Machine Learning (ML) services in a cloud environment presents a challenge to the cloud vendor with respect to the cloud container configuration sizing for any given customer use case. OracleLabs has developed an automated framework that uses nested-loop Monte Carlo simulation to autonomously scale any size customer ML use cases across the range of cloud CPU-GPU "Shapes" (configurations of CPUs and/or GPUs in Cloud containers available to end customers). Moreover, the OracleLabs and NVIDIA authors have collaborated on a ML benchmark study which analyzes the compute cost and GPU acceleration of any ML prognostic algorithm and assesses the reduction of compute cost in a cloud container comprising conventional CPUs and NVIDIA GPUs.

## I. INTRODUCTION

Cloud containers have seen increased use in the business world nowadays since they provide a separation of concerns, as end customers focus on their application logic and dependencies, while cloud vendors can focus on deployment, configuration, and security without bothering with application details. The technology enables businesses to access software on the internet as a service [1]. Cloud containers scale with the computing needs of the business, provide a high degree of customization, and reduce the Operations & Infrastructure costs for end customers (versus the huge overhead cost for customers operating their own datacenters). More importantly, major cloud vendors including Oracle, Google, Microsoft and Amazon charge cloud container services based on the specific use cases, number of users, storage space and compute costs across CPUs and GPUs in the customers' cloud tenancies. Hence, a company porting applications to a cloud environment will only pay for the services procured and choose a package that suits the customer's budget.

One challenge for deploying big-data ML services in a cloud environment wherein bare-metal containers and/or virtual machines (VMs) are populated with various "shapes" of CPUs and/or GPUs, is the appropriate container sizing. For prognostic ML applications with time-series sensor data (the focus of this paper), customer use cases vary all over the map, from a simple use case for monitoring one machine with 10 sensors and slow sampling rates, to huge Oil-and-Gas-size use cases with hundreds of thousands of high sampling rate sensors. In general, for any given customer engagement it would take a lot of trial-and-error runs by end customers guided by consultants with the cloud provider to discover optimal cloud configurations, which can vary enormously from customer-to-customer. Ideally, it would be nice to let a customer start small and autonomously grow their cloud container capabilities through "elasticity" as compute dynamics dictate. However, in practice that flexibility is not as smooth as cloud marketing teams might wish. The relationships between configuration resources (Memory, CPUs, GPUs) and cost estimates for ML use cases is not a simple "feeds and speeds" lookup table, because the compute cost for advanced ML prognostics use cases generally scale linearly with the number of observations (determined by sensor sampling rates), but (highly) nonlinearly with the number of sensors and the size of training dataset desired for training the ML algorithm. There is a steep nonlinear tradeoff between desired prognostic accuracy versus memory footprint and overhead compute cost. An example below illustrates a typical customer use case scenario of ML prognostic implemented in a cloud container:

1) Customer A has a use case with only 20 signals, sampled at a slow rate of just once per hour, such that a typical year's worth of data is a couple of MB.

2) Customer B has a fleet of Airbus 320's, each with 75000 sensors onboard, sampled at once per second, such that every plane generates 20 TB of data per month.

3) All other customers fall somewhere in the very wide use case range between A and B.

What is needed is a realistic way of pre-assessing, or "scoping" the cloud capability specifications for those two extreme use cases (1) and (2), so that the end customer and the cloud provider are able to scope out the cloud containers that would be the most appropriate reference for any prospective use cases (3).

OracleLabs has developed an Autonomous Cloud-Node Scoping Framework that fulfills an important function for customers interested in migrating ML applications from their on-prem data centers into cloud containers. We extend our prior work [2] and present here a Monte Carlo based scoping tool/technique for Oracle cloud containers consisting of CPUs and NVIDIA GPUs, for automatic evaluation of the compute cost of any ML algorithm as a parametric function of number of signals, number of observations, and number of desired training vectors (denoted as "three conventional ML design parameters" in the rest of the paper). Note there are many classes of ML algorithmics used for "classification". This paper deals instead with an important class of "prognostic" ML pattern recognition defined as nonlinear nonparametric regression, used for anomaly discovery in big-data dense-sensor IoT streaming analytics and time-series databases. The preferred ML service used in this paper is Oracle's advanced pattern recognition technique, the Multivariate State Estimation Technique (MSET2) [3-5], but the framework can accommodate other forms of pluggable prognostic ML techniques, including neural nets and support vector machines.

For this investigation, OracleLabs and our collaborators at NVIDIA have devised an automated ML compute cost benchmarking between CPUs and NVIDIA GPUs, which systematically and parametrically evaluates the compute cost of any ML algorithm and empirically assesses the non-linear relationships between the intensity of ML workload (customer use cases) and compute cost. We have performed a comprehensive compute cost evaluation and a GPU-speedup-factor evaluation for prospective end-customer use cases, ranging from tiny applications with 10s of sensors with slow sampling rates, to truly Big Data use cases involving terabytes of data per month from large fleets of assets. As such, the compute cost scoping framework presented herein benefits big data prognostic use cases for dense-sensor internet-of-things (IoT) use cases in such fields as Utilities, Oil and Gas, smart manufacturing, commercial aviation, and of course data center IT assets.

The remainder of this paper is organized as follows. Section II presents the implementation of the ContainerStress Framework, Oracle's MSET2 technique served as a pluggable ML technique, testing signals and the implementation of GPU algorithms. Section III.A introduces the 3D compute cost contours using CPU measured by the ContainerStress framework, and Section III.B illustrates the GPU accelerations of MSET2 over CPU executions. Section IV provides the conclusions.

## II. METHODOLOGY

### A. ContainerStress Framework Implmenetations

The ContainerStress autonomous scoping framework assesses the compute costs of any prognostic ML technique employed in the cloud container. This is achieved through a Monte Carlo based simulation as a parametric function of the three important ML scoping parameters. The goal is to perform setup/scoping estimation on the cloud container for different end customer use cases while adapting to any ML techniques in the category of statistical pattern recognition called nonlinear nonparametric regression. The output shows the computational overhead cost using 3D response-surface methodology (examples illustrated later) in terms of the compute cost for the conventional training process and streaming surveillance in ML techniques. Figure 1 illustrates the concept of the ContainerStress framework.

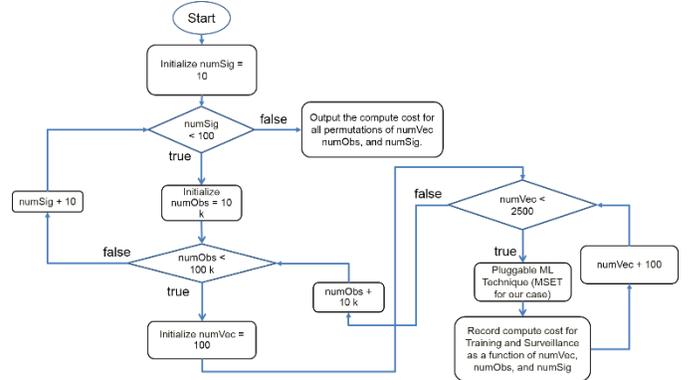

*Figure 1: Flowchart of ContainerStress framework for autonomous cloud container scoping.*

### B. Pluggable ML Prognostic Alogirhtm

The case study presented herein demonstrates the ContainerStress framework using MSET2, which is being used for prognostic surveillance of time series sensor signals for predictive maintenance applications. MSET2 provides very high sensitivity for proactive warnings of incipient anomalies, and ultra-low false-alarm and missed-alarm probabilities. Although this paper focuses on the performance of MSET2, we have architected ContainerStress to support pluggable ML algorithms so that other conventional forms of ML services such as Neural Nets, Support Vector Machines, Auto Associative Kernel Regression, which will also be easily evaluated in terms of cloud container configuration resources to meet different cloud customers' requirements.

### C. Realistic Testing Signals

The time-series signals used in the case study have been synthesized with a high-fidelity signal synthesis algorithm from real time series signatures across a variety of IoT industrial use cases. These signals are synthesized, not simulated, which match real IoT sensor signals in all statistical characteristics important to ML prognostics, including serial correlation content, cross correlation between/among signals, and stochastic content (variance, skewness, kurtosis), as real IoT sensor signals. For the large scale database of synthesized signals used in this investigation, OracleLabs' Telemetry Parameter Synthesis System (TPSS) has been employed [7-9].

## D. GPU Platform Algorithm Implementations

GPU architectures differ from CPU ones mainly in the fact that GPUs work by leveraging massive fine-grained parallelism of the order of 10000 threads. A typical CPU has on the order of 10 threads and is better at performing coarse grained parallelism. The main challenge with porting an algorithm to a GPU platform is to extract fine-grained parallelism and also efficiently use the memory subsystem of the GPU. The computational routines for MSET2 were implemented for the GPU platform using the CUDA programming model [6] where individual threads are grouped into *blocks*, which are grouped into a *grid* as shown in Figure 2. Further, a group of 32 threads is called a *warp* and all threads in a warp can issue the same instruction in any given cycle.

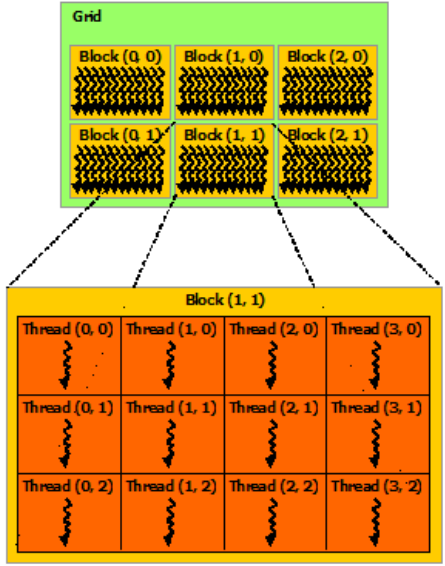

*Figure 2: Organization of threads in the CUDA programming model.*

The most computationally intensive part of MSET2 is the similarity matrix kernel that is a non-linear matrix binary operation. This computational routine was implemented in CUDA by decomposing the algorithm into logical hierarchical decompositions corresponding to an individual block, warp and thread. Correspondingly, the right memory channels for each level of decomposition is used to yield the highest performance possible. As in the case for a matrix multiplication, the compute cost of the similarity matrix increases faster than the amount of memory accesses. Hence, close attention is paid to efficient reuse of memory as well.

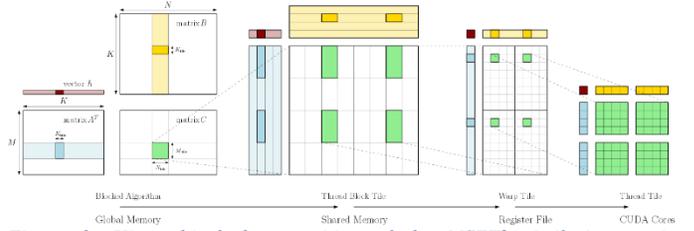

*Figure 3: Hierarchical decomposition of the MSET2 similarity matrix algorithm into kernel level, block level, warp level and thread level computations.*

Apart from the similarity matrix routine, MSET2 requires other routines like matrix multiplication and Eigen decomposition (Figure 3). For these, the cuBLAS [1] and cuSOLVER [2] libraries from NVIDIA are used. Custom kernels were written for all other routines required for the MSET2 algorithm.

## III. EVALUATION AND DISCUSSIONS

### A. Three Dimentional Compute Cost Contours

We demonstrate here the power and utility of our ContainerStress framework incorporating MSET2 in an Oracle cloud container and examine how compute cost varies with respect to the three ML parameters wherein MSET2 is employed as a cloud service. The compute cost measurements are presented by bars and the parametric cost function is represented by 3D response-surfaces, showing the real compute cost measurements and the observed trending to scope out the cloud implementation of MSET2 for prototypic advanced prognostic anomaly discovery applications with dense-sensor IoT industrial applications.

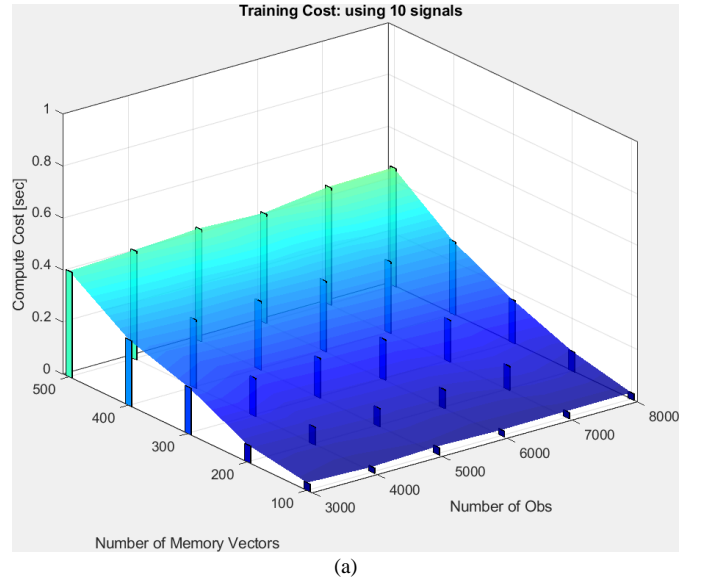

(a)

---

[1] https://developer.nvidia.com/cublas

[2] https://developer.nvidia.com/cusolver

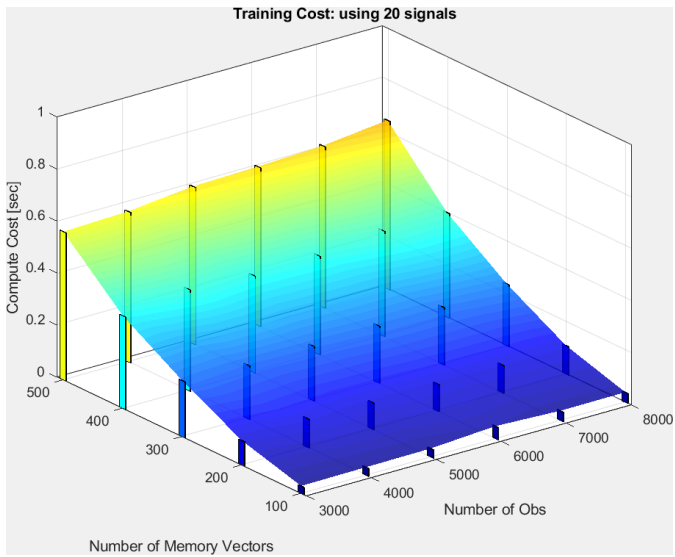

(b)

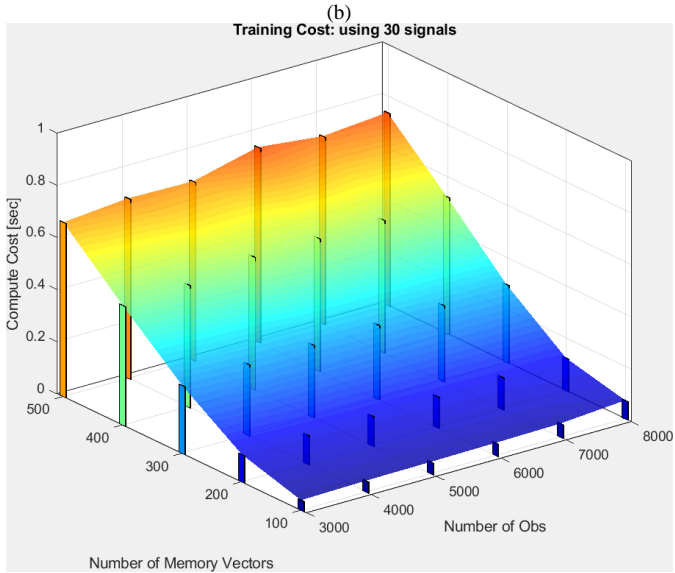

(c)

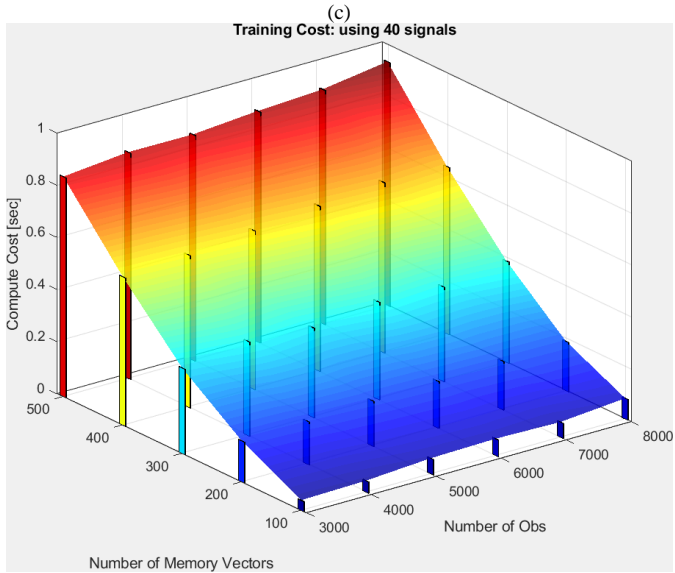

(d)

*Figure 4: The 3D compute cost contours of cloud implementation of MSET2 versus the number of memory vector, number of observations during Training process, and the number of signals is incremented by 10 at a time from (a) to (d). The blue and red color schemes represent the smallest and highest compute costs respectively.*

Figure 4 illustrates the parametric empirical relationships between compute cost and the three ML parameters in the Training process of MSET2. It can be concluded that the compute cost of Training process primary depends very sensitively on the number of memory vectors and number of signals.

Similarly, Figure 5 (a)-(d) illustrate the parametric empirical relationships between compute cost and the three ML parameters for streaming surveillance process. It can be observed that the compute cost of streaming surveillance primary depends on the number of observations and signals.

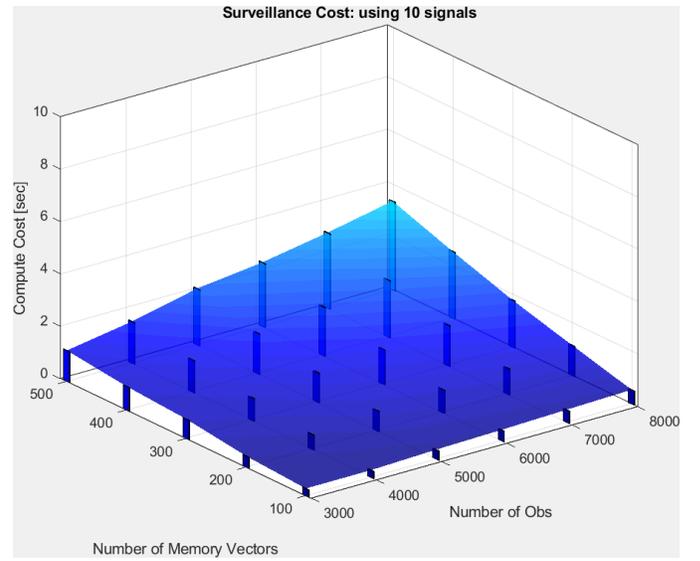

(a)

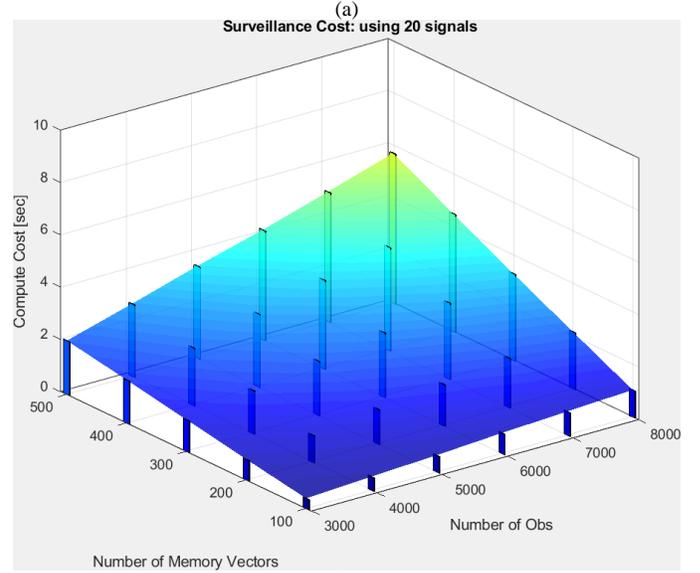

(b)

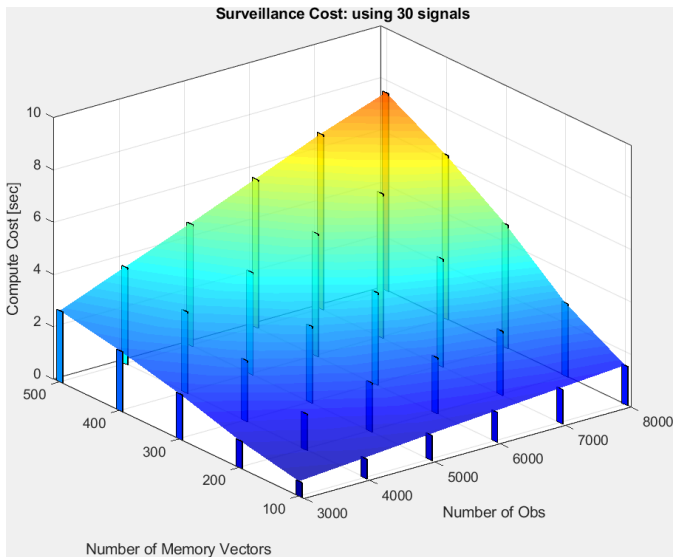

(c)

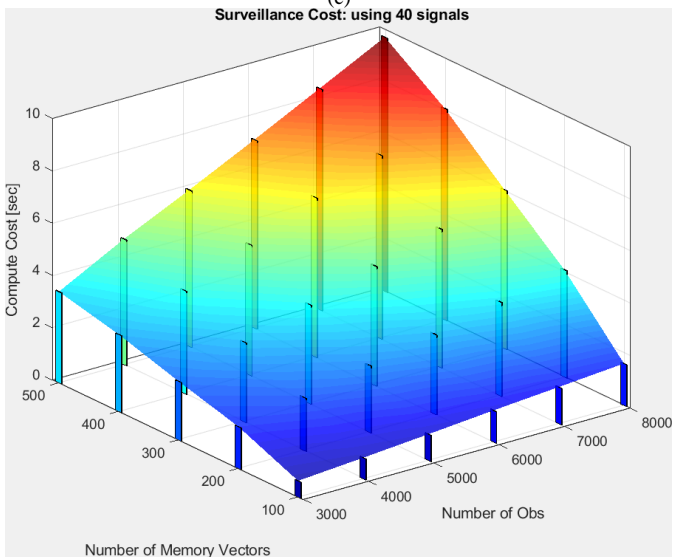

(d)

*Figure 5: The 3D compute cost contours of cloud implementation of MSET2 versus the number of memory vector, number of observations during surveillance streaming process, and the number of signals is incremented by 10 at a time from (a) to (d). The blue and red color schemes represent the smallest and highest compute costs respectively.*

With the 3D response-surface of compute cost above, we are able to quickly and efficiently scope out the appropriate configurations of the cloud container(s) for big-data customer applications with MSET as a service integrated for any given customer use cases.

### B. GPU Speedup Factor

In addition, we also deployed the ContainerStress framework on an Oracle cloud container, on which both CPU and GPU implementation of MSET2 were executed and benchmarked. One outstanding discovery made during the course of this investigation is the tremendous speedup factors (which is defined as the ratio of the compute cost for CPU-only and CPU+GPU cloud configurations) that are attained for any Oracle Cloud containers/VMs containing one or more NVIDIA GPUs for big-data ML use cases. Figures 6-8 show *measured* compute cost and GPU speedup-factors for a broad range of ML prognostic use cases with the latest CPUs (Intel Xeon Platinum) and NVIDIA GPUs (Tesla V100), where we have evaluated overhead compute costs and GPU speedup factors parametrically as a function of the three ML parameters. The relative influences of each ML parameter on the compute cost and attainable GPU speedup factors are also thoroughly investigated. Specifically, Figure 6 presents the speedup factor starts from 200x and can reach up to 1500x in the training process when number of signals varies from $2^5$ to $2^{10}$ and number of memory vectors varies from $2^7$ to $2^{13}$. Note that the missing parts in the training surface result from the ML training constraint that the number of memory vectors is at least twice the number of signals required by MSET2. Hence outputs are included only for these use cases meeting this required training constraint.

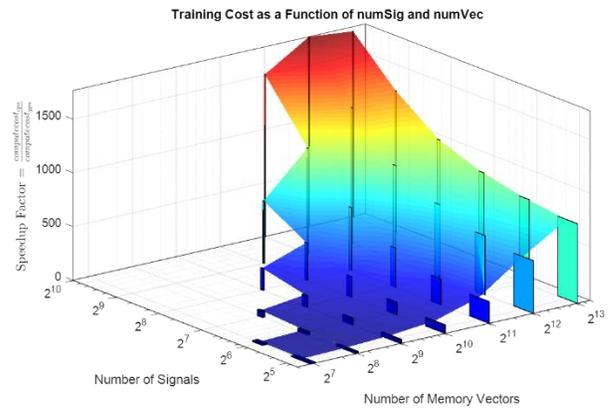

*Figure 6: The 3D training cost contours in term of speed factor as a function of number of signals and number of memory vectors. The X, Y axis are in log scale. The blue and red color schemes represent the smallest and highest compute costs respectively.*

Figure 7 illustrates the speedup factor in the surveillance part of MSET2 as a function of number of observations and number of memory vectors for the prognostic use cases consisting of 64 signals. It can be observed that even with a small IoT use case, the speedup factor still grows non-linearly and can exceed 5000x during the surveillance streaming process.

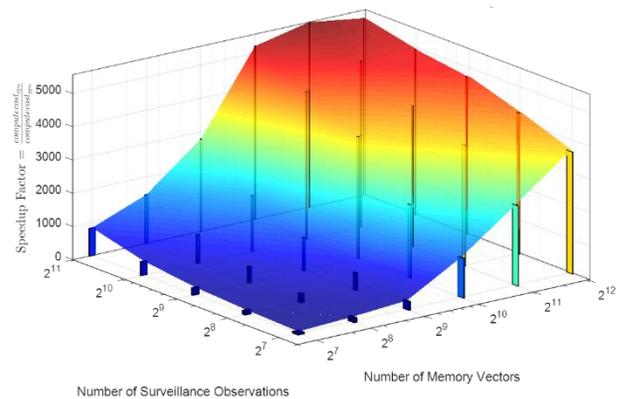

*Figure 7: The 3D surveillance cost contours in term of speed factor as a function of number of observations and number of memory vectors for 64-signal*

*use case. The X, Y axis are in log scale. The blue and red color schemes represent the smallest and highest compute costs respectively.*

Similarly, Figure 8 illustrates the speedup factor in the surveillance portion of MSET2 for the prognostic use cases consisting of 1024 signals. It can be concluded that with a larger IoT use case, the speedup factor further increases and can exceed 9000x during the surveillance streaming process.

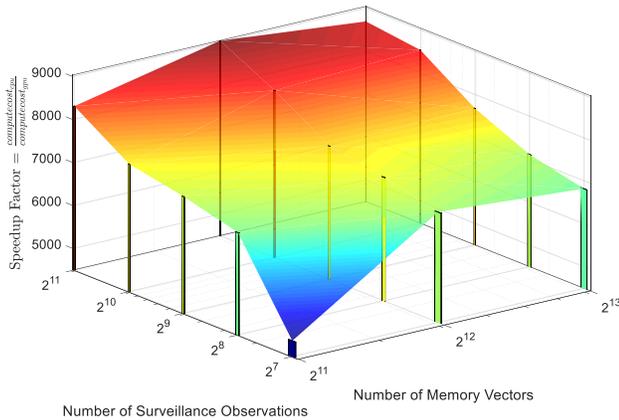

*Figure 8: The 3D surveillance cost contours in term of speed factor as a function of number of observations and number of memory vectors for 1024-signal use case. The X, Y axis are in log scale. The blue and red color schemes represent the smallest and highest compute costs respectively.*

In sum, the new ContainerStress framework incorporating NVIDIA GPU acceleration provides a robust and highly scalable approach for evaluating the deployability of a given ML prognostic technique in cloud containers/VMs comprising mixes of CPUs and NVIDIA GPUs. While we showcase in this paper that for a simple ML prognostic use case consisting of just 64 signals, a speedup factor of up to 1500 and 5000 is attained respectively for the training and the surveillance streaming process, the compute cost reduction is expected to be even much greater for larger scale use cases.

## IV. CONCLUSION

Advanced statistical ML algorithms are being developed, trained, tuned, optimized, and validated in a cloud environment for dense-sensor IoT prognostics applications in the fields of Oil-and-Gas, manufacturing, transportation (including aviation), utilities, and datacenters. The present challenge with offering prognostic ML pattern recognition in a cloud environment is sizing the customer container appropriately to ensure the customer has good performance, high throughputs, and low latencies, for real time streaming prognostics. OracleLabs has developed an autonomous cloud configuration-scoping framework called ContainerStress, which systematically evaluates the compute cost and the GPU acceleration factors, for a given ML technique as a parametric function of number of signals, number of observations, and number of training vectors, for scalable streaming prognostics in a cloud environment, and displays the compute cost results in with the aid of 3D response-surface methodology. In addition OracleLabs and NVIDIA have demonstrated the substantial acceleration power (upwards of 200x) on Oracle's advanced machine learning pattern recognition technique by using NVIDIA GPUs. This work will enable customers in dense-sensor IoT industries to harness vast amounts of data from sensors, processes, and physical assets to gain valuable prognostic insights and to proactively terminate or avoid system degradation events that could challenge overall asset availability goals or diminish safety margins for life-critical industrial settings, and, when Oracle's MSET2 is the ML algorithm employed, achieve the foregoing prognostic goals with ultra-low false-alarm probabilities.